\documentclass[twocolumn,prb,showpacs]{revtex4}

\usepackage{amssymb}
\usepackage{latexsym}
\usepackage{graphics}
\usepackage{color}
\usepackage{graphicx}

\begin{document}

\title{Vortex states in 2D superconductor at high magnetic field in a periodic
pinning potential. }
\author{V.Zhuravlev$^{1}$ and T.Maniv$^{1,2}$}
\email{maniv@tx.technion.ac.il}
\affiliation{$^{1}$Chemistry Department, Technion-Israel Institute of Technology, Haifa 32000, Israel}
\affiliation{$^{2}$Grenoble High Magnetic Field Laboratory, Max-Planck-Institute f\"{u}r Festkorperforschung and Center National de la Recherche Scientific,25 Avenue des Martyres, F-38042, Cedex 9, FRANCE }
\date{\today}

\begin{abstract}
The effect of a periodic pinning array on the vortex
state in a 2D superconductor at low temperatures is studied within
the framework of the Ginzburg-Landau approach. It is shown that
attractive interaction of vortex cores to a commensurate pin
lattice stabilizes vortex solid phases with long range positional
order against violent shear fluctuations. Exploiting a simple
analytical method, based on the Landau orbitals description, we
derive a rather detailed picture of the low temperatures vortex
state phase diagram. It is predicted that for sufficiently clean
samples application of an artificial periodic pinning array would
enable one to directly detect the intrinsic shear stiffness anisotropy
characterizing the ideal vortex lattice.
\end{abstract}
\pacs{ 74.20.De, 74.25.Dw, 74.25.Qt}

\maketitle

\section{Introduction}

The nature of the vortex lattice melting transition in 2D superconductors
has been debated in the literature for many years \cite{mzvwrmp}. Early
proposals \cite{doniahub}$^{,}$\cite{fisher80}, based on the similarity to the
Kosterlitz-Thouless-Halperin-Nelson-Young theory of melting in 2D solids\cite
{kthny}, have led to the conclusion that the melting transition is
continuous. A weak first order melting transition was predicted more
recently, however, by several Monte Carlo simulations \cite{tesanovic}$^{,}$\cite
{humacdon} using the Ginzburg-Landau (GL) theory. It has been shown recently
\cite{mzvwrmp}$^{,}$\cite{zm99}$^{,}$\cite{zm02} that shear motions of Bragg chains
along the principal crystallographic axis of the vortex lattice cost a very
small fraction of the SC condensation energy and are responsible for the low
temperature vortex lattice melting.

This intrinsic anisotropy of the vortex lattice with respect to
shear stress can not be easily detected experimentally since the
orientation of the principal axis with respect to the laboratory
frame depends on the local pinning potential, which in real
superconductors is usually produced by random distribution of
pinning centers. Indirect experimental detection of this hidden
anisotropy may be achieved by means of the small angle neutron
scattering (SANS) technique, due to the 1D nature of the effective
thermal fluctuations in the vortex liquid state just above the
melting point (see Ref.\cite{zm02}). A direct detection of this
anisotropy (e.g. by means of SANS) could be possible if vortex
solid phases with long range positional order were stabilized
against the random influence of pinning impurities. \ This can be
achieved by exposing the SC sample to an artificial periodic
pinning array and tuning the magnetic flux density to an integer
multiple of the pinning centers density. As will be shown in this
paper, under certain conditions the artificial periodic pinning
potential can stabilize weakly pinned vortex solid
phases with long range positional order, which may exhibit the
shear stiffness anisotropy characterizing the ideal vortex
lattice.

Vortex matter interacting with periodic pinning arrays is currently a
subject of intense experimental \cite{fiory78}$^{-}$\cite{terentiev00} and
theoretical \cite{nelson79}$^{-}$\cite{pogosov02} investigations. Developments of
nano-engineering techniques, such as e-beam lithography, make it possible to
fabricate well defined periodic arrays of sub-micron antidotes, or magnetic
dots, in SC films with low intrinsic pinning, enabling to study the effect
of well controlled artificial pinning centers. These experiments have shown
that under certain conditions the underlying artificial pinning centers can
attract vortices very strongly, thus stabilizing vortex patterns with global
translational symmetry against the randomize influence of the natural
pinning centers.

From theoretical point of view the utilization of an external periodic
pinning potential provides a convenient tool for testing different models of
the vortex state by simplifying considerably the model calculations. At the
same time, however, the interplay between the vortex-vortex interactions,
which favor hexagonal vortex lattice symmetry, and the underlying periodic
potential can lead to a variety of vortex configurations, depending on the
pinning strength, in which vortices detach from pinning centers to form more
closely packed vortex patterns.

As the interaction with a periodic substrate stabilizes the vortex
system versus thermal fluctuations, it generally increases the
melting temperature. However, as we shall see in this paper,
deviation from the ideal hexagonal symmetry due to pinning reduces
the phase dependent interaction between vortex chains \cite{zm02},
making them less enduring under thermal fluctuations. In the weak
pinning limit, where depinned floating state can occur, the
corresponding phase diagram becomes rather complicated, due to the
possibility of transitions between floating solid and pin solid
phases \cite{reichhardt01}.

In the present paper we study the influence of a periodic pinning substrate
on the vortex state in 2D , extreme type II superconductors, at
perpendicular high magnetic fields. Our approach is based on the previously
developed theory of vortex lattice melting in pure superconductors (\cite{mzvwrmp}$^{,}$\cite
{zm99}$^{,}$\cite{zm02}), carried out within the framework of the
GL theory in the lowest Landau level (LLL) approximation. Specializing the
calculation for a vortex system interacting with a square pinning array
under the first matching magnetic field, we study in detail some key
limiting regions of the vortex phase diagram, which enables us to determine
its main qualitative features.

\section{The model}

\label{sec:1}

We consider a 2D superconductor at high perpendicular magnetic field,
interacting with a periodic substrate of pinning centers, located at $\left(
x_{i},y_{j}\right) $. A phenomenological Ginzburg-Landau functional, with an
order parameter $\psi \left( x,y\right) $, is used to describe the
superconducting (SC) part of the free energy, and a local periodic pinning
potential \cite{tesanovic94}:
\begin{equation}
V_{pin}=v_{0}\sum_{i,j}|\psi \left( x_{i},y_{j}\right) |^{2}  \label{Vpin}
\end{equation}
determines the interaction of the vortex state with pinning centers. We assume $%
v_{0}>0$, so that the pinning energy is minimal if the vortex core
positions, determined by $\psi \left( x_{i},y_{j}\right) =0$, coincide with
pinning centers.

Our main interest here is in the influence of the pinning potential on the
vortex lattice melting process, so that the pinning energy $V_{pin}$ is
restricted to the range of the vortex lattice melting energy, which is much
smaller than the SC condensation energy. Since the latter is of the same
magnitude as the cyclotron energy, it is justified to restrict the analysis
to the LLL of the corresponding SC order parameter, which can be therefore
written as a linear combination of ground Landau orbitals:
\begin{eqnarray}
\psi \left( x,y\right) &=&\sum_{n}c_{n}\phi _{q_{n}}\left( x,y\right)
\label{LOrep} \\
c_{n}=|c_{n}|e^{i\varphi _{n}}; &\hspace{0.5cm}&\phi _{q}\left( x,y\right)
=e^{2iqx-\left( y+q\right) ^{2}}  \nonumber
\end{eqnarray}
where $q_{n}=qn$, $q=\pi /a_{x}$, and the amplitudes $c_{n}$ in the mean
field approximation are related to the (spatial) mean square SC order
parameter, $\Delta _{0}^{2}$ , through: $|c_{n}|^{2}=c_{0}^{2}=\sqrt{\frac{%
2q^{2}}{\pi }}\Delta _{0}^{2}$. In our notations all space variables are
measured in units of magnetic length.

\begin{figure}
\begin{center}
\includegraphics[width=6cm]{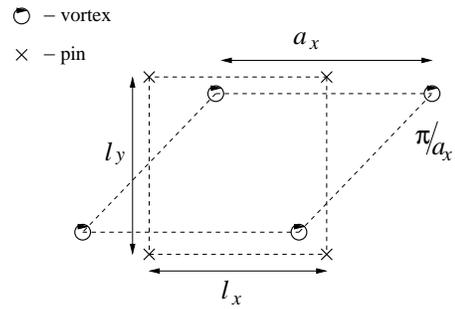}
\end{center}
\caption{Schematic arrangement of a
vortex lattice relative to the pin lattice.}
\label{fig:0}
\end{figure}

In this model, due to the Gaussian attenuation along the $y$-axis over a
characteristic distance of the order of the magnetic length, the vortex
cores (located at the zeros of $\psi \left( x,y\right) $ ) form a network of
linear chains along the $x$-axis, each of which is determined mainly by a
superposition of two neighboring Landau orbitals \cite{zm02}. The
parameter $a_{x}$ is therefore\ equal to the inter-vortex distance within a
chain, while $\pi /a_{x}$ is the inter-chain spacing in the $y$-direction
(see Fig. \ref{fig:0}). It should be noted that deviations of Landau orbital
(LO) amplitudes from their mean field value, $c_{0}$, resulting in strong
local distortions of the superfluid density and large increase of the
corresponding free energy density, are neglected in comparison with
variations of the phase variables, $\varphi _{n}$, provided the orbital
direction is selected to be along the principal crystallographic axis \cite
{zm99}$^{,}$\cite{zm02}.

We select the pinning centers to form a rectangular lattice
\begin{equation}
\left( x_{i},y_{j}\right) =\left( l_{x}i+x_{0},l_{y}j+y_{0}\right)
\label{eq:sqlat}
\end{equation}
where $i,j=0,\pm 1,....$. The parameters $x_{0}$ and $y_{0}$ determine the
relative position of the pin and vortex lattices. The nature of the vortex
state in the presence of the pinning potential depends crucially on the
ratio of the number of vortices, $N=\sqrt{N}\times \sqrt{N}$, to the number
of pinning centers, $N_{p}=N_{p,x}\times N_{p,y}$ . Since the density of
vortices depends on the external magnetic field strength $H$, one can tune
this ratio by varying $H$. Of special interest are the matching fields $%
H=H_{\nu }$ , $\nu =1,2,...$, when the ratio $n_{\phi }\equiv N/N_{p}=\nu $
is an integer.

In matching fields one may distinguish between two different situations,
when vortices are bound or unbound to pinning centers. If the pin lattice
and the vortex lattice unit cells are commensurate along both $x$ and $y$
directions , i.e. $l_{x}=c_{x}a_{x}$ , $l_{y}=c_{y}\pi /a_{x}$, with $c_{x}$
and $c_{y}$ being integers, the pinning energy is equal to zero, since all
the vortices coincide with pinning centers. In all other cases of matching
fields , $c_{x}c_{y}=\pi \times {\mathtt{integer}}$ , non of the numbers \ $%
c_{x}\geq 1$ and $c_{y}\geq 1$ can be integer , and the lattice constants
are incommensurate in both directions. It will be shown below that such a
vortex configuration is in a floating state with respect to the pin lattice,
similar to vortex states in mismatching magnetic fields.

Using the LO representation, Eq.(\ref{LOrep}), of the SC order parameter in
Eq.(\ref{Vpin}) for the pinning energy, one may take advantage of the
localized nature of the LOs and expand $V_{pin}$ in the small parameter $%
\lambda =e^{-q^{2}}$, which reflects the small overlap integral between
adjacent orbitals contributing to the local superfluid density at the
pinning centers. Retaining only dominant terms in $\lambda $ Eq. (\ref{Vpin}%
) is reduced to the form:
\begin{eqnarray}
V_{pin} &=&V_{0}\sqrt{\frac{2q^{2}}{\pi }}\sum_{k}\left[ u_{k}+2%
\sum_{m=1}e^{-q^{2}m^{2}/2}u_{k+m/2}\Phi _{k,m}\right]  \nonumber \\
&\simeq &V_{0}\sqrt{\frac{2q^{2}}{\pi }}\sum_{k}\left[
u_{k}+2e^{-q^{2}/2}u_{k+1/2}\Phi _{k,1}\right]  \nonumber \\
u_{k+m/2} &=&\frac{1}{N_{p,y}}\sum_{j}e^{-2\left( y_{j}+q\left( k+m/2\right)
\right) ^{2}}  \nonumber \\
\Phi _{k,m} &=&\frac{1}{N_{p,x}}\sum_{i}\cos \left( \varphi _{k+m}-\varphi
_{k}+2mqx_{i}\right)  \label{eq:3}
\end{eqnarray}
where $V_{0}=v_{0}\Delta _{0}^{2}N_{p,x}N_{p,y}$. It should be stressed that
this approximation is valid only for LOs along the principal axes since the
minimal distance between them, $q=\pi /a_{x}$, is sufficiently large to
ensure small and rapidly decreasing value of the overlap integrals between
more distant orbitals.

If $c_{x}$ is not an integer, namely the rectangular pin lattice and vortex
lattice are incommensurate in $x$-direction, then Eq.(\ref{eq:3}) shows that
$\Phi _{k,m}=0$. In this case the pinning energy does not depend on the
phases (i.e. the relative horizontal positions ) of the Landau orbitals.

Expressing the functions $u_{k}$ and $u_{k+1/2}$ with the help of Poisson
summation formula as
\begin{eqnarray}
&&u_{k+m/2}=\frac{1}{N_{p,y}}\sum_{j}e^{-2\left( l_{y}j+y_{0}+q\left(
k+.5\right) \right) ^{2}}\approx \frac{1}{N_{p,y}}\sqrt{\frac{\pi }{%
2l_{y}^{2}}}\times  \nonumber \\
&&\left[ 1+2\sum_{j}e^{-\frac{\pi ^{2}j^{2}}{2l_{y}^{2}}}\cos \left( \frac{%
2\pi j\left( q(k+\frac{m}{2})+y_{0}\right) }{l_{y}}\right) +...\right]
\label{eq:6}
\end{eqnarray}
we note that when the lattices are incommensurate also along the $y$ axis (
i.e. when both $c_{x}$ and $c_{y}$ are not integer) the oscillating terms in
$u_{k}$ are averaged to zero after summation over $k$. Thus, the pinning
energy for incommensurate lattices is a constant
\begin{equation}
V_{pin}=V_{0}\frac{q\sqrt{N}}{l_{y}N_{p,y}}=V_{0}
\end{equation}
which does not depend on the mutual orientation of the vortex and the pin
lattices. Note that the system size in $y$ direction is $L_{y}=q\sqrt{N}%
=l_{y}N_{p,y}$ , a relation connecting $\sqrt{N}$ to $N_{p,y}$. Obtained
result is valid only for large system, $N\longrightarrow \infty $, where the
boundary effect can be neglected.

For the sake of simplicity, we will consider in what follows a square pin
lattice with $n_{\phi }=1$. In the commensurate situation the pinning energy
is minimal ( i.e. equal to zero) when all vortices coincide with pinning
centers. Deviations of vortices from this configuration in the form of shear
distortions along the principal crystallographic axes are of special
interest due to the relatively low SC energy involved. For the principal
axis parallel to a side of the square unit cell, $c_{x}=c_{y}=1$ and $%
q^{2}=\pi $, and so, according to Eq. (\ref{eq:3}), the pinning energy per
single vortex is, up to small terms of the order $\sim e^{-2\pi }$, given
by:
\begin{eqnarray}
\frac{V_{pin}}{N} &=&v\frac{1}{\sqrt{N}}\sum_{k}\left[ a_{1}-a_{2}\cos
\left( \varphi _{k}-\varphi _{k-1}\right) \right]  \nonumber \\
&\simeq &\kappa _{x}v\frac{1}{\sqrt{N}}\sum_{k}\left[ 1-\cos \left( \xi
_{k}\right) \right]  \label{eq:15}
\end{eqnarray}
where $v=V_{0}/N$ , $a_{1}=1-2e^{-\pi /2}\simeq .584$ and $a_{2}=2e^{-\pi
/2}\left( 1+2e^{-\pi /2}\right) \simeq .589$ . Note that in the above
expressions we set $x_{0}=y_{0}=q/2$ so that the minimal pinning energy is
obtained for $\xi _{k}=0$. Note also that for the undistorted square lattice
in which $\varphi _{k}=\varphi _{k-1}$, the expression in the first line of
Eq. (\ref{eq:15}) is not strictly zero since $a_{1}\neq a_{2}$. \ The error,
which is of order higher than the second in $e^{-\pi }$ , can be neglected
in the approximation leading to Eq. (\ref{eq:15}). The numbers $a_{1},a_{2}$
can be thus considered equal within this approximation, allowing us to
introduce a single coefficient $\kappa _{x}\equiv a_{1}\simeq a_{2}\simeq
.59 $. The expression in the second line of Eq. (\ref{eq:15}) yields the
correct (i.e. zero) value for the undistorted lattice. It is written in
terms of the variables, $\xi _{k}\equiv \varphi _{k}-\varphi _{k-1}$,
describing the lateral positions of the vortex chains, which are generated
mainly by interference between two neighboring LOs. This is consistent with
the well known definition $u_{x}=\partial \varphi /\partial y$ of vortex
displacement along the $x$ axis in the long wavelength limit \cite{Moore89}.

\begin{figure}
\begin{center}
\includegraphics[width=6cm]{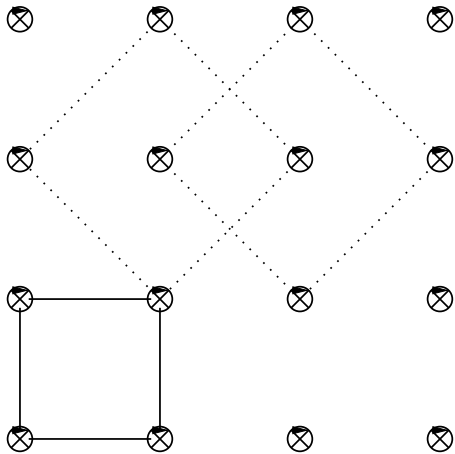}
\end{center}
\caption{Primitive and none-primitive unit cell representations
(solid and dotted lines respectively) of the square pin lattice
used for description of shear distortion along the principal axes
of the vortex lattice.} \label{fig:2n}
\end{figure}

To evaluate the excess pinning energy associated with shear distortion along
the diagonal of the square unit cell the pin lattice may be conveniently
described by two interpenetrating simple square sub-lattices with $%
c_{x}=1,c_{y}=2$ and $q^{2}=2\pi $ ( see Fig.\ref{fig:2n} ). The
corresponding interchain pinning energy for each of the sublattices can
again be obtained from Eq. (\ref{eq:15}), with $\kappa _{x^{\prime }}\simeq
.84$ and a phase shift of $\pi /2$ , which arises due to different shape of
the unit cell.

The SC part of the free energy functional for the commensurate lattices
described above ($a_{x}=\sqrt{\pi }$) is given by the following ($\xi _{k}$%
-dependent) expression \cite{zm99}:
\begin{equation}
\frac{H_{sc}}{N}=-h_{\Box }-T_{\Box }\frac{1}{\sqrt{N}}\sum_{k}\left[ 1-\cos
(\xi _{k+1}-\xi _{k})\right]  \label{eq:sqlaten}
\end{equation}
where $h_{\Box }$\ is the SC condensation energy (per unit flux) of the
square vortex lattice, and $T_{\Box }=\frac{4\lambda _{sq}^{2}}{1+4\lambda
_{sq}}h_{\Box }$ is the shear distortion energy parameter, expressed through
the dimensionless interchain coupling constant, $\lambda _{sq}=\exp (-\pi )$%
. \ Here $h_{\Box }=\varepsilon _{0}\frac{\beta _{A}}{\beta _{sq}}$ , where $%
\beta _{A}\simeq 1.159$ and $\beta _{sq}\simeq 1.18$ , are the values of the
Abrikosov structure parameter for regular hexagonal and square lattices
respectively, and $\varepsilon _{0}$ is the SC condensation energy of the
former.

\begin{figure}
\begin{center}
\includegraphics[width=6cm]{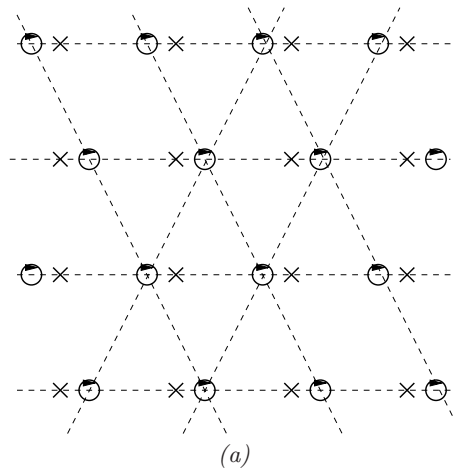}\\
\textit{(a)}\\
\vspace{.5cm} \includegraphics[width=6cm]{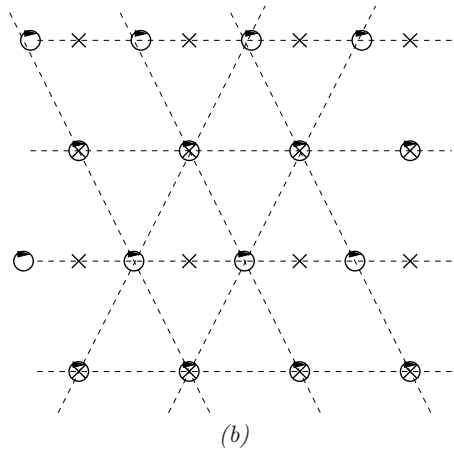}\\
\textit{(b)}
\end{center}
\caption{(a) The vortex lattice state with the lowest energy,
which is commensurate with a square pinning lattice in the limit
of zero pinning strength. (b) An alternative vortex lattice state,
which may be favorable under square framework boundary conditions
(see text).} \label{fig:3n}
\end{figure}

For the specific choice $\xi _{k}=\gamma k$, where $\gamma $ is a
constant, the Bragg family of vortex chains along the principal
axis,denoted $x$, is characterized by a lateral displacement, $\xi
_{k+1}-\xi _{k}=\gamma $, between neighboring chains. Evidently,
the SC energy, $H_{sc}$ , for the
undistorted square vortex lattice $\xi _{k}=0$ ( $\gamma =0$ ) ( see Eq. (%
\ref{eq:sqlaten}) ), is equal to $Nh_{\Box }$. However, the minimum of the
SC energy with respect to the collective tilt angle parameter $\gamma $ is
reached for a triangular vortex lattice, determined by $\xi _{k}=\pi k$ ($%
\gamma =\pi $), whose unit cell is an isosceles triangle with a base ( along
$x$-axis) and a height equal to $\sqrt{\pi }$ (see Fig. \ref{fig:3n}). The
corresponding SC energy is equal to $H_{sc}/N=-h_{\Box }-2T_{\Box }$. This
value is lower than the SC energy of the square vortex lattice, and only
slightly higher ( i.e. by $\sim .45\%$) than the SC energy of the
equilateral triangular (Abrikosov) lattice, $H_{\triangle }/N=-\varepsilon
_{0}$.

\section{Vortex states for the lowest matching field}

\subsection{ Commensurate and incommensurate Ground states}
\label{sec:3.1}

The competition between the pinning energy, Eq. (\ref{eq:15}), which favors
vortices approaching the pinning points on a square lattice, and the SC
energy, Eq. (\ref{eq:sqlaten}), preferring triangular lattice configuration,
leads to 'frustrated' vortex structures, which depend on the relative
pinning strength.

At zero temperature they can be obtained by minimizing the total energy,
consisting of the SC and pinning parts. Since in the LO representation each
orbital is $\sqrt{N}$-fold degenerate, the effective Hamiltonian is written
as:
\begin{eqnarray}
f_{\Box }&=&\frac{H_{sc}+V_{pin}}{N}=\frac{T_{\Box }}{\sqrt{N}}\sum_{k} \Big\{
-h_{\Box }/T_{\Box } \nonumber \\
&+&4p\left[ 1-\cos (\xi _{k})\right] -\left[ 1-\cos (\xi _{k+1}-\xi _{k})\right]
 \Big\}\label{eq:low_ph}
\end{eqnarray}
where the parameter $p\equiv \kappa _{x}v_{\Box }/4T_{\Box }$ determines the
strength of the pinning potential relative to the inter vortex chain
coupling. Under the constraints imposed by the requirement of
commensurability between the vortex configuration and the pin lattice, the
vortex chains are restricted to move laterally along the common $x$-axis of
the underlying lattices (see Fig. \ref{fig:0}). The corresponding
displacements, $\xi _{k}$ , may be separated into two groups, corresponding
to even and odd vortex chains, as follows:
\begin{equation}
\xi _{k}=\left\{
\begin{array}{lll}
\theta _{l} & \textrm{for} & k=2l \\
\zeta _{l} & \textrm{for} & k=2l-1
\end{array}
\right.
\end{equation}
so that
\begin{eqnarray}
f_{\Box }&=&\frac{T_{\Box }}{\sqrt{N}}\sum_{l}\Big\{ -2h_{\Box }/T_{\Box }+
4p\left[ 2-\cos (\theta _{l})-\cos (\zeta _{l})\right]  \nonumber \\
&-&\left[ 2-\cos (\theta _{l}-\zeta _{l})-\cos (\theta _{l-1}-\zeta _{l})\right]
\Big\}  \label{eq:low_ph1}
\end{eqnarray}

The calculation may be greatly simplified if we assume that the stationary
point values within each group are all equal, that is: $\theta _{l}=\theta
_{c}$ and $\zeta _{l}=\zeta _{c}$. \ This restriction may be justified in
the weak pinning regime $0<p<1$ , where the dominant SC energy part favors
periodic triangular vortex structures, as shown in Fig. \ref{fig:3n}a.

Substituting these values to Eq. (\ref{eq:low_ph1}) and minimizing the
resulting functional one finds that the total energy has a minimum:
\begin{equation}
\min \left( f_{\Box }\right) =-\frac{1}{\sqrt{N}}\sum_{k}\left[ 2T_{\Box
}(1-p)^{2}+h_{\Box }\right]  \label{eq:low_ph2}
\end{equation}
at $\theta _{c}=-\zeta _{c}=\theta _{0}$ where
\begin{equation}
\cos (\theta _{0})=p.  \label{eq:gr_st}
\end{equation}
Thus, at zero pinning strength, $p=0$, the ground state energy per unit
flux, $f_{\Box }=-2T_{\Box }-h_{\Box }$, corresponds to a triangular vortex
lattice configuration, $\theta _{c}=-\zeta _{c}=\pi /2$ , whereas in the
opposite extreme, when $p=1$ , the ground state energy, $f_{\Box }=h_{\Box }$%
, corresponds to a square vortex lattice, $\theta _{c}=-\zeta _{c}=0$ ,
which coincides with the underlying pin lattice. It should be stressed,
however, that due to the constraints imposed by the requirement of
commensurability with the pin square lattice, the triangular vortex
structure obtained in the zero pinning strength limit, is not the
equilateral (Abrikosov) lattice (see Fig.\ref{fig:3n}a). \ This
discontinuity indicates that the transition to the depinned (floating)
vortex lattice should be of the first order (see below).

An illustration of the weak pinning ground state configuration is shown in
Fig. \ref{fig:3n}a. It is seen that odd and even vortex chains are shifted
in opposite directions symmetrically with respect to the underlying
substrate. The relative positions of the two lattices are determined by the
strength of the pinning potential, $V_{0}$. In the zero pinning limit, $%
V_{0}\rightarrow 0$, the vortices in odd (even) chains approach lattice
points which are shifted laterally by a quarter of a lattice constant, $%
\frac{1}{4}l_{x}$ ( $l_{x}=\sqrt{\pi }$ ), in the positive (negative) sense
with respect to the square pin lattice, forming isosceles triangular
lattice. Note that the asymmetric configuration, shown in Fig.\ref{fig:3n}b,
in which half of the vortex chains remain pinned to the underlying
substrate, has energy $-2T_{\Box }(1-2p)-h_{\Box }$, which is only slightly
(i.e. by a small, second order correction in $p$ ) higher than the energy
given by Eq. (\ref{eq:low_ph2}). Such an asymmetric configuration may become
energetically favorable (see Ref. \cite{reichhardt01}$^{,}$\cite{pogosov02}) due
to, e.g. boundary conditions which are incompatible with the even-odd chain
symmetry described above.

At sufficiently weak pinning, when the pinning energy becomes comparable to
the difference between the SC energies of the commensurate isosceles
triangular vortex lattice with $a_{x}^{2}=\pi $ , and the incommensurate
equilateral triangular lattice with $a_{x}^{2}=\sqrt{3}\pi /2$ (Abrikosov
lattice), the latter is preferable. To show this note that the energy, $%
-h_{\triangle }$ , of the equilateral triangular vortex lattice in the
presence of incommensurate pin lattice is influenced only by the average
pinning potential $v$, so that:
\begin{equation}
-h_{\triangle }=-\varepsilon _{0}+v=-\varepsilon _{0}+\frac{4}{\kappa _{x}}%
pT_{\Box }
\end{equation}
Comparing this value with that obtained in Eq. (\ref{eq:low_ph2}) for the
commensurate, isosceles triangular lattice, $-2T_{\Box }(1-p)^{2}-h_{\Box }$%
, we find that for $p\leq $ $p_{c}\simeq .25$ the floating equilateral
triangular lattice is the lowest energy state. This critical point can be
thus identified as a transition point from pinned (commensurate) solid to a
floating (incommensurate) solid state.

A second critical point exists in the strong pinning regime, i.e. at $p=1$,
as indicated by Eq.(\ref{eq:gr_st}), which has no real solution at any $p>1$%
. \ At this critical point the vortex lattice coincides with the square pin
lattice ( i.e. $\theta _{0}=0$ in Eq.(\ref{eq:gr_st}))\ and the pinning
energy reaches its absolute minimum value (i.e. zero ). Since any further
increase of the pinning strength above the critical value, $p=1$, can not be
compensated by the SC energy terms in Eq.(\ref{eq:low_ph1}), the vortex
configuration remains fixed at the square lattice structure for any $p\geq 1$%
. Thus, with increasing values of the parameter $p$ , the ground state
vortex configuration changes continuously from a triangular
lattice at $p=p_{c}$ , into a square lattice at $p=1$ , which does not
changes with further increase of the pinning strength. This continuous
transformation from a triangular lattice to a square lattice can be
classified as a second order phase transition at $p=1$.

\subsection{Commensurate equilibrium states at finite temperature}

\label{sec:4}

In the ideal vortex state at finite temperature thermal fluctuations
associated with the low-lying shear excitations along the principal
crystallographic axis destroy the long range phase coherence of the vortex
state and lead to melting of the ideal vortex lattice at a temperature, $%
T_{m}$, well below the mean field $T_{c}$. \ This feature indicates an
intrinsic anisotropy of the ideal vortex crystal \cite{zm02}: The
characteristic excitation energy for sliding vortex chains along the
principal axis ( denoted by $x$ ) parallel to a side of the unit cell is two
orders of magnitude smaller than the SC condensation energy, and one order
of magnitude smaller for fluctuations along the diagonal axis ( denoted by $%
x^{\prime }$). For all other crystallographic orientations the shear energy
is of the order of the SC condensation energy.

\begin{figure}
\begin{center}
\includegraphics[width=8cm]{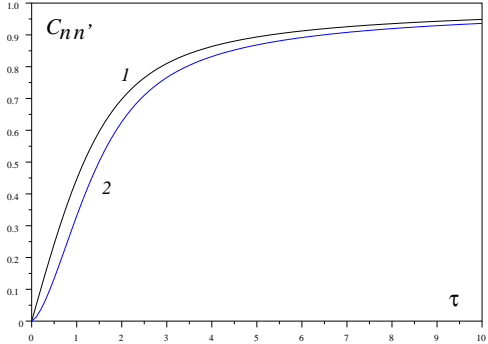}
\end{center}
\caption{Pair correlation function between nearest Landau
orbitals: (1) at strong pinning, as a function of the
dimensionless inverse temperature $\protect\tau =4pT_{\square
}/T$, and (2) in the triangular Abrikosov
vortex lattice with a single pinned chain , $n=0$, as a function of $\widetilde{%
\protect\tau }\equiv T_{\triangle }/T$}.  \label{fig:4}
\end{figure}

The nucleation of a SC crystallite can be established in such an ideal model
by selecting boundary conditions which fix the position of a single vortex
chain with respect to the laboratory frame. As shear fluctuations of
parallel vortex chains diverge with the distance from the fixed chain \cite{zm99}, 
a SC domain is restricted to nucleate only around a pinned chain,
its transverse size shrinking to that of a single magnetic length as the
temperature rises toward $T_{m}$. \ \ For the sake of simplicity, we avoid
here the complication associated with the discontinuous nature of the vortex
lattice melting process, which involves two principal families of easily
sliding Bragg chains \cite{mzvwrmp}, and restrict the analysis to a single
family of vortex chains, i.e. that with the lowest crossover temperature $%
T_{cm}$ \cite{zm02}. \ A meaningful definition of $T_{cm}\left( a_{x}\right)
$ may invoke the phase correlation function,
\begin{equation}
C_{n^{\prime },n}\equiv \langle e^{i\left( \varphi _{n^{\prime }}-\varphi
_{n}\right) }\rangle  \label{Cn'n}
\end{equation}
between Landau orbitals, $\ n^{\prime }$ and $n\ $,$\ $located near the
fixed chain $n=0$ . $\ $Thus, melting of the entire vortex lattice occurs
essentially when phase correlation between the nearest neighboring chains (
i.e. $\ n=1$, $n^{\prime }=2$ ) closest to the fixed chain is significantly
suppressed (e.g. by a factor of 1/2). \ In the $p\rightarrow 0$ limit we use
the expression derived in Ref.(\cite{zm99}) to find: \
\[
C_{n^{\prime }=2,n=1}\left( \widetilde{\tau }\right) \simeq \left( \frac{%
I_{1}(\widetilde{\tau })}{I_{0}(\widetilde{\tau })}\right) \left( \frac{%
I_{1/2}(\widetilde{\tau })}{I_{0}(\widetilde{\tau })}\right)
\]
with $\widetilde{\tau }\equiv T_{\triangle }/T$ . \ Here the characteristic
temperature, $T_{\triangle }\simeq \frac{4\lambda ^{2}}{1+4\lambda }%
\varepsilon _{0}$ with $\lambda =\exp (-\sqrt{3}\pi /2)$, corresponds to
interaction between the principal LOs in the equilateral triangular lattice
state. Note that the crossover between the vortex solid state at zero
temperature, where $\widetilde{\tau }\rightarrow \infty $ , and $%
C_{n^{\prime }=2,n=1}\left( \widetilde{\tau }\right) \rightarrow 1$ , and
the vortex liquid state at high temperature, where $\widetilde{\tau }%
\rightarrow 0$ , and $C_{n^{\prime }=2,n=1}\left( \widetilde{\tau }\right)
\rightarrow 0$ , occurs at about $\widetilde{\tau }\simeq 1.5$ , so that $%
T_{cm}\left( a_{x}\right) \approx .67T_{\triangle }$ (see Fig. \ref{fig:4}%
). \ This crossover temperature is close to , though somewhat lower than
the melting temperature, $T_{m}\simeq 1.2T_{\triangle }\simeq 2.8T_{\Box }$,
predicted in Ref.( \cite{zm99}).

The presence of the periodic pinning potential stabilizes the vortex lattice
against the violent phase fluctuations discussed above. This effect is
nicely demonstrated by the phase correlation function $C_{n^{\prime },n}$ (
Eq.(\ref{Cn'n}) ), which controls the mean superfluid density (see Eq.(\ref
{LOrep})) near the melting point. Assuming strong pinning, $p\gg 1$, and
neglecting the small GL inter-vortex-chain coupling, the correlation
function can be determined from the expression:
\begin{equation}
C_{n^{\prime },n}\simeq \frac{\prod_{k}\int_{0}^{\pi }d\xi _{k}e^{i\left(
\varphi _{n^{\prime }}-\varphi _{n}\right) }e^{-4p\tau \cos (\xi _{k})}}{%
\prod_{k}\int_{0}^{\pi }d\xi _{k}e^{-4p\tau \cos (\xi _{k})}}
\label{Cn'nPin}
\end{equation}

Using the identity $\varphi _{n}=\sum_{k=n_{0}}^{n}\xi _{k}$, where the
value of $n_{0}$ can be found from boundary conditions which influence only
the global phase of the SC order parameter, we find that
\begin{eqnarray}
C_{n^{\prime },n} &\simeq &\left( \frac{I_{1}(4p\tau )}{I_{0}(4p\tau )}%
\right) ^{|\Delta n|}  \nonumber \\
&\simeq &\exp \left( -|\Delta n|/8p\tau \right) \hspace{0.2cm}\textrm{for}%
\hspace{0.2cm}4p\tau \gg 1,  \label{pin}
\end{eqnarray}
where $\Delta n\equiv n^{\prime }-n$. This result contrasts with the
correlation function obtained in the pure state \cite{zm99}, which has the
asymptotic form
\begin{equation}
C_{n^{\prime },n}\propto \exp \left( -\frac{\overline{n}}{2\tau }|\Delta
n|^{2}\right) \quad\textrm{for} \quad\tau \gg 1  \label{depin}
\end{equation}
where $\overline{n}=\frac{n^{\prime }}{3}+\frac{2n}{3}-\frac{1}{2}$ , with
the $n=0$ chain being fixed. As discussed above (see also Refs.(\cite{zm99}%
$^{,}$\cite{mzvwrmp}$^{,}$\cite{zm02})), fixing chain positions through boundary
conditions is physically equivalent to introducing pinning potential into
the GL functional, which is a crucial step for stabilizing the vortex
lattice. The global stability of the vortex lattice in the presence of the
periodic pinning potential is reflected in Eq.(\ref{pin}), as compared with
Eq.(\ref{depin}), by the translational invariance of the former correlation
function, as well as by its relatively weak (simple exponential) decay. \ \

To determine the crossover temperature from the square pin solid (SPS) to
the vortex liquid we may follow the procedure described above and find the
temperature $T_{cm}\left( a_{x},p\right) $ at which $C_{n^{\prime },n}$ in
Eq.(\ref{Cn'nPin}) for $|\Delta n|=1$ is reduced by a factor of 1/2 with
respect to its zero temperature ($\tau \rightarrow \infty $ ) limit. \ This
yields in the strong pinning limit, $p\gg 1$ (see Fig. \ref{fig:4}), the
linear dependence
\begin{equation}
T_{cm}\left( a_{x},p\right) \approx .86\times 4pT_{\Box }  \label{T_cm(p)}
\end{equation}

Beside its influence on the vortex lattice melting transition, the pinning
potential can change the vortex lattice structure, both continuously and
discontinuously. The zero temperature limit was discussed in Sec.(\ref{sec:3.1}).
Above the critical value $p=1$ the lateral vortex positions coincide with
the pin square lattice positions, $\xi _{l}=0$. For decreasing pinning
strength below $p=1$, the configuration of the vortex lattice deviates
continuously from the square structure to a lattice with
vortices shifted along chains away from the pinning centers.

Similar second order SPS to TPS phase transition (as a function of $p$ ) is
expected at finite temperatures. Indeed, as shown in Sec.(\ref{sec:3.1}), the free
energy functional $f_{\Box }$ in Eq. (\ref{eq:low_ph}) is minimized at the
stationary points $\xi _{k}=\xi _{ck}\equiv \left( -1\right) ^{k}\theta _{0}$
, with $\cos \theta _{0}=p$ for $p\leq 1$ , and at $\xi _{ck}=0$ for $p\geq
1 $. \ Thus, expanding $f_{\Box }$ as a Taylor series in $\left( \xi
_{k}-\xi _{ck}\right) $ about its stationary points it is clear that for $%
p\geq 1$ (when $\xi _{ck}=0$ ) the expansion includes only \textit{even}
powers $\xi _{k}$ ( due to the symmetry of $f_{\Box }$ with respect to $\xi
_{k}\longrightarrow -\xi _{k}$ ). Thus, at any finite temperature, $T$ , the
thermal mean values $\left\langle \xi _{k}\right\rangle $ are equal to zero
for $p\geq 1$ , implying that for pinning strengths above the critical value
$p=1$ the mean vortex positions coincide with the square pin lattice.

\section{Phase diagram for the lowest matching field}

\label{sec:5}

\begin{figure}
\begin{center}
\includegraphics[width=10cm]{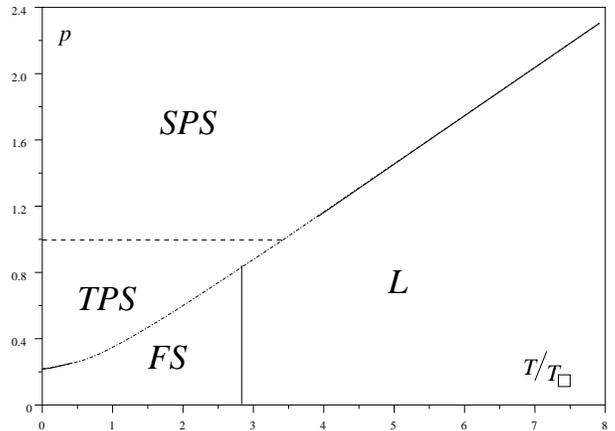}
\end{center}
\caption{$V_{0}-T$ phase diagram: Solid lines - first order phase
transitions from SPS to L phase (at large pinning strength), from
TPS to FS (near zero temperature), and between FS and L phases.
Dashed line - second order phase transition between partially
pinned (TPS) and fully pinned (SPS) vortex crystals. The
dashed-dotted line connects smoothly between the asymptotic SPS-L
line and the low temperatures TPS-FS line (see text for
explanation). } \label{fig:ph_diag}
\end{figure}

The results of the previous sections enable us to draw a rather clear
picture of the $V_{0}-T$ -phase diagram, as shown in Fig. \ref{fig:ph_diag}.
In the strong pinning regime, $p\gg 1$, the pinning strength is so large
that the gain in commensurate energy is larger than the vortex-vortex energy
gain at any temperature, and so the floating solid phase is not favorable.
Thus, the vortex lattice melting in this region should take place directly
from the SPS to the liquid phase, as described by the asymptotic expression,
Eq.(\ref{T_cm(p)}), which is equivalent to the straight line $p\approx
0.29T/T_{\Box }$ in the large $p$ regime of the phase diagram.

In the small $p$ regime the stable phase at low temperatures is the FS. Here
the energy gain associated with creation of the closed packed equilateral
triangular vortex lattice, exceeds the energy cost of the incommensurate
state. This state remains stable up to a relatively high temperature $%
T\simeq 2.8T_{\Box }$ , above which it melts into a vortex liquid state. The
phase boundary in this region is vertical (i.e. independent of $p$ ) since
it is determined by the vortex-vortex coupling and not by the pinning energy
(which is a constant in the floating state).

In the low temperatures region of the phase diagram our analysis shows the
existence of two phase transitions: \ At small pinning, increasing $p$ above
$p_{c}\approx .25$ transforms the FS discontinuously to a pin solid since
the energy gain associated with the commensurate pin vortex solid exceeds
the energy cost of distorting the closed packed equilateral triangular
vortex lattice. The discontinuous nature of this transition is due to the
fact that even infinitesimal deviation from a commesurate configuration
rises the pinning energy by a finite amount (i.e. at least from $.6v$ to $v$%
).

It turns out that the pin vortex crystal just above the
commensurate-incommensurate transition is not a square lattice, as found by
Reichhardt \textit{et al}. \cite{reichhardt01}, but a triangular one, with a
unit cell which depends on the pinning strength.\ At $T=0$ it is a
parallelogram with equal base and height, which transforms continuously to a
square at $p=1$. \ Similar continuous transition from a frustrated
triangular pin lattice to the SPS \ takes place at the critical pinning
strength $p=1$ at any temperature $T$. \ Interestingly, the corresponding
horizontal transition line intersects the extrapolated SPS-L boundary line
at $T\approx \allowbreak 3.\,\allowbreak 44T_{\Box }$, $p=1$ , that is in
the close vicinity of the intersection between the vertical FS melting line,
$T\simeq 2.8T_{\Box }$, and the SPS-TPS line.

It is not exactly known, however, how the FS-TPS boundary is extended beyond
the zero temperature region. It is conceivable that its high temperature
sector coincides with the low temperature sector of the SPS melting line.
This is due to the fact that, at a fixed value of $p$, the driving force for
both transitions are thermal fluctuations involving sliding vortex chains,
which suppress the pinning energy gained in the commensurate phase (i.e. the
term $-4p\cos (\xi _{k})$ in Eq.(\ref{eq:low_ph})). In the SPS-L
transitions, where the vortex-vortex interaction is relatively small, this
suppression leads to uncorrelated vortex chains, resulting in melting. \ In
the TPS-FS transitions, where the the vortex-vortex coupling is relatively
large, the suppression of the pinning energy results in mutually correlated
vortex chains, which lose correlation with the underlying pinning lattice.

An intermediate pin solid phase of a triangular form has been also found in
the London model calculation reported by Pogosov \textit{et al}. \cite
{pogosov02}. However, in contrast to the Ginzburg-Landau model, discussed
here, they predicted the vortex configurations shown in Fig. \ref{fig:3n}b\
as preferable below some critical value of the pinning potential strength.
Above this value the symmetry of vortex lattice is changed discontinuously
to the square symmetry of the pin lattice.\

Our proposed phase diagram, shown in Fig.\ref{fig:ph_diag}, thus consists
of 2 pin solid phases, a floating solid and a liquid phase, delimitted by 4
interphase boundary lines, which intersect at two nearby triple points. This
result is similar to the phase diagram found by Reichhardt \textit{et al}.
\cite{reichhardt01}, using molecular dynamics simulations. \ However, the
intermediate TPS phase obtained in our calculation, is missing in Reichhardt
\textit{et al. }This seems to be due to the square boundary conditions
imposed in the latter calculation. Another difference concerns the zero
temperature limit of the PS-FS line, which seems to approach $p=0$ in
Reichhardt \textit{et al}..

\section{Conclusions}

The influence of a periodic pinning potential on the vortex state
of a 2D superconductor at temperatures well below the mean field
$T_{c}$ has been studied within the framework of the GL functional
integral approach. It is shown that attractive interaction of
vortex cores to a commensurate pin lattice stabilizes vortex solid
phases with long range positional order against violent shear
fluctuations along the principal crystallographic axis. Exploiting
a simple analytical approach we draw a rather detailed picture of
the relevant vortex state $p-T$ (pinning strength-temperature)
phase diagram. In agreement with previous numerical simulations
\cite {reichhardt01}, we have found a pinned, commensurate solid
phase in the strong pinning-low temperature part of the phase
diagram, which melts into a vortex liquid at high temperatures,
and transforms into a floating (incommensurate) solid at low
temperatures. \ We have shown that at low temperature, similar to
Ref. \cite{pogosov02}, there is an intermediate triangular phase,
where vortices detaching  from pinning centers remain strongly
correlated with them. This pinned (frustrated) triangular solid
transforms continuously into the fully pinned (square) solid phase
at $p=1$, and discontinuously to a floating solid at small pinning
strengths. The zero temperature limit of this
commensurate-incommensurate transition line occurs at a finite
pinning strength ($p=p_{c}\approx .25$).

It is predicted that for sufficiently clean samples, where random
pinning is weak enough, application of an artificial periodic
pinning array with an appropriate strength would stabilize a
weakly pinned vortex solid phase with long range
positional order. Exploiting the SANS method to the sample under
these conditions one could therefore directly detect the shear
stiffness anisotropy characterizing the ideal vortex lattice.

\begin{acknowledgments}
This research was supported in parts by a grant from the Israel Science
Foundation founded by the Academy of Sciences and Humanities, and by the
fund from the promotion of research at the Technion. 
\end{acknowledgments}

\end{document}